\documentclass[3p,times,12pt]{elsarticle}
\usepackage[english]{babel}
\usepackage{float}
\usepackage{graphicx}
\usepackage{amssymb}
\usepackage{amsfonts}
\usepackage{amsmath}
\usepackage{amsthm}
\usepackage[utf8]{inputenc}
\usepackage[normalem]{ulem}
\usepackage{empheq,etoolbox}
\usepackage{hyperref}

\usepackage{tikz}
\usetikzlibrary{patterns}
\usetikzlibrary{calc}
\usepackage{amssymb}

\def\ri{2}
\def\rii{3}
\def\riii{4.0}
\def\riv{4.2}
\def\zi{8}
\def\zii{10}

\def\NameA{Cooler}
\def\NameB{Insulator}
\def\NameC{Heater}
\def\NameD{Insulator}
\def\NameE{Terminal}

\def\NameR{$r_0^\ast$}
\def\NameRi{$r_1^\ast$}
\def\NameRii{$r_2^\ast$}
\def\NameRmax{$r_{\max}$}
\def\NameZ{$z_0$}
\def\NameZmax{$z_{\max}$}

\newcommand{\midlabelline}[3]{
	\node (midlabel) at ($ (#1)!.5!(#2) $) {#3};
	\draw[-, ultra thin] (#1) --  (midlabel);
	\draw[-, ultra thin] (midlabel) -- (#2);
}

\newcommand{\quartlabelline}[3]{
	\node (midlabel) at ($ (#1)!.65!(#2) $) {#3};
	\draw[-, ultra thin] (#1) --  (midlabel);
	\draw[-, ultra thin] (midlabel) -- (#2);
}

\newcommand{\toplabelline}[3]{
	\node (midlabel) at ($ (#1)!.85!(#2) $) {#3};
	\draw[-, ultra thin] (#1) --  (midlabel);
	\draw[-, ultra thin] (midlabel) -- (#2);
}

\def\injectorA#1#2{%
#1
\if\relax\detokenize{#2}\relax
\begin{tikzpicture}
\else
\begin{tikzpicture}[scale=#2]
\fi

\node at (0,0)[circle,fill,inner sep=1pt]{};

\draw[pattern=north west lines] (0,0) rectangle (\zii,\ri);
\draw[pattern=horizontal lines] (0,\ri) rectangle (\zi,\rii);
\draw[pattern=dots] (0,\rii) rectangle (\zi,\riii);
\draw[pattern=vertical lines] (0,\riii) rectangle (\zi,\riv);
\draw[pattern=crosshatch] (\zii,0) rectangle (\zii+0.2,\riv);

\draw[ultra thin] (0,0)--(-1.65,0);
\draw[ultra thin] (-1.65,\riv)--(0,\riv);
\toplabelline{-1.6,0}{-1.6,\riv}{\NameRmax}
\draw[ultra thin] (-1.25,\riii)--(0,\riii);
\midlabelline{-1.2,0}{-1.2,\riii}{\NameRii}
\draw[ultra thin] (-0.85,\rii)--(0,\rii);
\midlabelline{-0.8,0}{-0.8,\rii}{\NameRi}
\draw[ultra thin] (-0.45,\ri)--(0,\ri);
\midlabelline{-0.4,0}{-0.4,\ri}{\NameR}
\midlabelline{0,-0.4}{\zi,-0.4}{\NameZ}
\midlabelline{0,-0.8}{\zii,-0.8}{\NameZmax}
\draw[ultra thin] (0,0)--(0,-0.85);
\draw [ultra thin, densely dashed] (\zi,\ri) -- (\zi, 0);
\draw[ultra thin] (\zi,0)--(\zi,-0.45);
\draw[ultra thin] (\zii,0)--(\zii,-0.85);

\draw[line width=0.3mm] (9,1.8) -- (9.5,2.8) node[pos=1.3] {\normalsize 0};
\draw[line width=0.3mm] (7.5,2.5) -- (8.5,2.7) node[pos=1.3] {\normalsize 1};
\draw[line width=0.3mm] (7.5,3.5) -- (8.5,3.7) node[pos=1.3] {\normalsize 2};
\draw[line width=0.3mm] (7.5,4.1) -- (8.5,4.3) node[pos=1.3] {\normalsize 3};
\draw[line width=0.3mm] (10.1,2.5) -- (11.1,3.5) node[pos=1.3] {\normalsize 4};
\end{tikzpicture}}

\def\dBx{8}
\def\dBy{6}

\def\injectorB#1#2{%
#1
\if\relax\detokenize{#2}\relax
\begin{tikzpicture}
\else
\begin{tikzpicture}[scale=#2]
\fi

\node at (0,0)[circle,fill,inner sep=1pt]{};
\node at (\dBx,\dBy)[circle,fill,inner sep=1pt]{};

\draw[ultra thin] (\dBx,0.3+\dBy)--(0,\riv);
\draw[ultra thin] (\dBx,\dBy)--(0,0);
\draw[ultra thin] (\dBx+5,\dBy)--(\zii,0);
\draw[ultra thin] (\dBx+4,0.3+\dBy)--(\zi,\riv);

\draw (0+\dBx,0+\dBy)--(5+\dBx,0+\dBy)--(5+\dBx,0.24+\dBy)--(4+\dBx,0.24+\dBy)--(4+\dBx,0.3+\dBy)
 --(0+\dBx,0.3+\dBy)--(0+\dBx,0+\dBy);
\draw[ultra thin] (-5+\dBx,0.24+\dBy)--(-4+\dBx,0.24+\dBy)--(-4+\dBx,0.3+\dBy)
  --(0+\dBx,0.3+\dBy);
\draw[ultra thin] (5+\dBx,-0.24+\dBy)--(4+\dBx,-0.24+\dBy)--(4+\dBx,-0.3+\dBy)
  --(0+\dBx,-0.3+\dBy);
\draw[ultra thin] (-5+\dBx,-0.24+\dBy)--(-4+\dBx,-0.24+\dBy)--(-4+\dBx,-0.3+\dBy)
  --(0+\dBx,-0.3+\dBy);
\draw[pattern=crosshatch] (5+\dBx,-0.4+\dBy) rectangle (\dBx+5.1,0.4+\dBy);
\draw[pattern=crosshatch] (-5+\dBx,-0.4+\dBy) rectangle (\dBx-5.1,0.4+\dBy);
\draw[ultra thin, dotted] (-5+\dBx,\dBy)--(\dBx,\dBy)--(\dBx,\dBy-0.3);

\draw[pattern=north west lines] (0,0) rectangle (\zii,\ri);
\draw[pattern=horizontal lines] (0,\ri) rectangle (\zi,\rii);
\draw[pattern=dots] (0,\rii) rectangle (\zi,\riii);
\draw[pattern=vertical lines] (0,\riii) rectangle (\zi,\riv);
\draw[pattern=crosshatch] (\zii,0) rectangle (\zii+0.2,\riv);
	
\draw[ultra thin] (0,0)--(-1.65,0);
\draw[ultra thin] (-1.65,\riv)--(0,\riv);
\toplabelline{-1.6,0}{-1.6,\riv}{\NameRmax}
\draw[ultra thin] (-1.25,\riii)--(0,\riii);
\midlabelline{-1.2,0}{-1.2,\riii}{\NameRii}
\draw[ultra thin] (-0.85,\rii)--(0,\rii);
\midlabelline{-0.8,0}{-0.8,\rii}{\NameRi}
\draw[ultra thin] (-0.45,\ri)--(0,\ri);
\midlabelline{-0.4,0}{-0.4,\ri}{\NameR}
\midlabelline{0,-0.4}{\zi,-0.4}{\NameZ}
\midlabelline{0,-0.8}{\zii,-0.8}{\NameZmax}
\draw[ultra thin] (0,0)--(0,-0.85);
\draw [ultra thin, densely dashed] (\zi,\ri) -- (\zi, 0);
\draw[ultra thin] (\zi,0)--(\zi,-0.45);
\draw[ultra thin] (\zii,0)--(\zii,-0.85);
	
\draw[line width=0.3mm] (9,1.8) -- (9.5,2.8) node[pos=1.3] {\normalsize 0};
\draw[line width=0.3mm] (7.5,2.5) -- (8.5,2.7) node[pos=1.3] {\normalsize 1};
\draw[line width=0.3mm] (7.5,3.5) -- (8.5,3.7) node[pos=1.3] {\normalsize 2};
\draw[line width=0.3mm] (7.5,4.1) -- (8.5,4.3) node[pos=1.3] {\normalsize 3};
\draw[line width=0.3mm] (10.1,2.5) -- (11.1,3.5) node[pos=1.3] {\normalsize 4};
\end{tikzpicture}}

\def\dCx{3}
\def\dCy{5.5}

\def\injectorC#1#2{%
#1
\if\relax\detokenize{#2}\relax
\begin{tikzpicture}
\else
\begin{tikzpicture}[scale=#2]
\fi

\node at (0,0)[circle,fill,inner sep=1pt]{};
\node at (\dCx,\dCy)[circle,fill,inner sep=1pt]{};

\draw [->, thick] (\dCx,\dCy)--(\dCx,\dCy+1);
\draw [->, thick] (\dCx,\dCy)--(\dCx+5.8,\dCy);
\draw (\dCx+5.8,\dCy) node [below] {$z$};
\draw (\dCx,\dCy+1) node [left] {$r$};

\begin{scope}
\clip(-10.8,0) rectangle (0,4.3);
\draw [pattern=vertical lines] (-6.5,0) circle (4.2);
\draw [fill=white] (-6.5,0) circle (4.0);
\draw [pattern=dots] (-6.5,0) circle (4.0);
\draw [fill=white] (-6.5,0) circle (3.0);
\draw [pattern=horizontal lines] (-6.5,0) circle (3.0);
\draw [fill=white] (-6.5,0) circle (2.0);
\draw [pattern=north west lines] (-6.5,0) circle (2.0);
\end{scope}
\node at (-6.5,0)[circle,fill,inner sep=1pt]{};

\draw[ultra thin] (\dCx,0.3+\dCy)--(0,\riv);
\draw[ultra thin] (\dCx,\dCy)--(0,0);
\draw[ultra thin] (\dCx+5,\dCy)--(\zii,0);
\draw[ultra thin] (\dCx+4,0.3+\dCy)--(\zi,\riv);
	
\draw (0+\dCx,0+\dCy)--(5+\dCx,0+\dCy)--(5+\dCx,0.24+\dCy)--(4+\dCx,0.24+\dCy)--(4+\dCx,0.3+\dCy)
  --(0+\dCx,0.3+\dCy)--(0+\dCx,0+\dCy);
\draw[ultra thin] (-5+\dCx,0.24+\dCy)--(-4+\dCx,0.24+\dCy)--(-4+\dCx,0.3+\dCy)
  --(0+\dCx,0.3+\dCy);
\draw[ultra thin] (5+\dCx,-0.24+\dCy)--(4+\dCx,-0.24+\dCy)--(4+\dCx,-0.3+\dCy)
  --(0+\dCx,-0.3+\dCy);
\draw[ultra thin] (-5+\dCx,-0.24+\dCy)--(-4+\dCx,-0.24+\dCy)--(-4+\dCx,-0.3+\dCy)
  --(0+\dCx,-0.3+\dCy);
\draw[pattern=crosshatch] (5+\dCx,-0.4+\dCy) rectangle (\dCx+5.1,0.4+\dCy);
\draw[pattern=crosshatch] (-5+\dCx,-0.4+\dCy) rectangle (\dCx-5.1,0.4+\dCy);
\draw[ultra thin, dotted] (-5+\dCx,\dCy)--(\dCx,\dCy)--(\dCx,\dCy-0.3);
	
\draw[pattern=north west lines] (0,0) rectangle (\zii,\ri);
\draw[pattern=horizontal lines] (0,\ri) rectangle (\zi,\rii);
\draw[pattern=dots] (0,\rii) rectangle (\zi,\riii);
\draw[pattern=vertical lines] (0,\riii) rectangle (\zi,\riv);
\draw[pattern=crosshatch] (\zii,0) rectangle (\zii+0.2,\riv);
	
\draw[ultra thin] (-6.5,\riii)--(0,\riii);
\draw[ultra thin] (-6.5,\rii)--(0,\rii);
\draw[ultra thin] (-6.5,\ri)--(0,\ri);

\draw[ultra thick, color=white] (-1.6,0)--(-1.6,\riv-0.05);
\draw[ultra thick, color=white] (-1.2,0)--(-1.2,\riii-0.05);
\draw[ultra thick, color=white] (-0.8,0)--(-0.8,\rii-0.05);
\draw[ultra thick, color=white] (-0.4,0)--(-0.4,\ri-0.05);

\draw[ultra thin] (0,0)--(-2.3,0);
\draw[ultra thin] (-6.5,\riv)--(0,\riv);

\toplabelline{-1.6,0}{-1.6,\riv}{\NameRmax}
\quartlabelline{-1.2,0}{-1.2,\riii}{\NameRii}
\midlabelline{-0.8,0}{-0.8,\rii}{\NameRi}
\midlabelline{-0.4,0}{-0.4,\ri}{\NameR}
\midlabelline{0,-0.4}{\zi,-0.4}{\NameZ}
\midlabelline{0,-0.8}{\zii,-0.8}{\NameZmax}
\draw[ultra thin] (0,0)--(0,-0.85);
\draw [ultra thin, densely dashed] (\zi,\ri) -- (\zi, 0);
\draw[ultra thin] (\zi,0)--(\zi,-0.45);
\draw[ultra thin] (\zii,0)--(\zii,-0.85);
	
\draw (0,0) node [below left] {(0,0)};

\draw[pattern=north west lines] (-10.7,-1.75) rectangle (-9.7,-1.25);
\draw (-9.7,-1.5) node [right] {\NameA};

\draw[pattern=horizontal lines] (-6.7,-1.75) rectangle (-5.7,-1.25);
\draw (-5.7,-1.5) node [right] {\NameB};

\draw[pattern=dots] (-2.7,-1.75) rectangle (-1.7,-1.25);
\draw (-1.7,-1.5) node [right] {\NameC};

\draw[pattern=vertical lines] (1.3,-1.75) rectangle (2.3,-1.25);
\draw (2.3,-1.5) node [right] {\NameD};

\draw[pattern=crosshatch] (5.3,-1.75) rectangle (6.3,-1.25);
\draw (6.3,-1.5) node [right] {\NameE};
\draw [ultra thin] (-10.7,0)--(-2.3,0);
\end{tikzpicture}}

\def\dxx{6.2}
\def\dyy{4}
\def\injectorD#1#2{%
#1
\if\relax\detokenize{#2}\relax
\begin{tikzpicture}
\else
\begin{tikzpicture}[scale=#2]
\fi

\node at (\dxx,\dyy)[circle,fill,inner sep=1pt]{};
\draw [->, thick] (\dxx,\dyy)--(\dxx,\dyy+1);
\draw [->, thick] (\dxx,\dyy)--(\dxx+5.8,\dyy);
\draw (\dxx+5.8,\dyy) node [below] {$z$};
\draw (\dxx,\dyy+1) node [left] {$r$};

\draw (0+\dxx,0+\dyy)--(5+\dxx,0+\dyy)--(5+\dxx,0.24+\dyy)--(4+\dxx,0.24+\dyy)--(4+\dxx,0.3+\dyy)
  --(0+\dxx,0.3+\dyy)--(0+\dxx,0+\dyy);
\draw[ultra thin] (-5+\dxx,0.24+\dyy)--(-4+\dxx,0.24+\dyy)--(-4+\dxx,0.3+\dyy)
  --(0+\dxx,0.3+\dyy);
\draw[ultra thin] (5+\dxx,-0.24+\dyy)--(4+\dxx,-0.24+\dyy)--(4+\dxx,-0.3+\dyy)
  --(0+\dxx,-0.3+\dyy);
\draw[ultra thin] (-5+\dxx,-0.24+\dyy)--(-4+\dxx,-0.24+\dyy)--(-4+\dxx,-0.3+\dyy)
  --(0+\dxx,-0.3+\dyy);
\draw[pattern=crosshatch] (5+\dxx,-0.4+\dyy) rectangle (\dxx+5.1,0.4+\dyy);
\draw[pattern=crosshatch] (-5+\dxx,-0.4+\dyy) rectangle (\dxx-5.1,0.4+\dyy);
\draw[ultra thin, dotted] (-5+\dxx,\dyy)--(\dxx,\dyy)--(\dxx,\dyy-0.3);
	
\draw[pattern=north west lines] (7,-0.15) rectangle (8,0.35);
\draw[pattern=horizontal lines] (7,0.45) rectangle (8,0.95);
\draw[pattern=dots] (7,1.05) rectangle (8,1.55);
\draw[pattern=vertical lines] (7,1.65) rectangle (8,2.15);
\draw[pattern=crosshatch] (7,2.25) rectangle (8,2.75);
\draw (8.1,0.1) node [right] {\NameA};
\draw (8.1,0.7) node [right] {\NameB};
\draw (8.1,1.3) node [right] {\NameC};
\draw (8.1,1.9) node [right] {\NameD};
\draw (8.1,2.5) node [right] {\NameE};
\midlabelline{0+\dxx,-0.5+\dyy}{\zi/2+\dxx,-0.5+\dyy}{\NameZ}
\midlabelline{0+\dxx,-0.9+\dyy}{\zii/2+\dxx,-0.9+\dyy}{\NameZmax}
\draw[ultra thin] (\dxx,\dyy)--(\dxx,\dyy-1);
\draw[ultra thin] (\dxx+4,\dyy-0.3)--(\dxx+4,\dyy-0.6);
\draw[ultra thin] (\dxx+5,\dyy)--(\dxx+5,\dyy-1);
\draw (\dxx-0.1,\dyy-0.4) rectangle (\dxx+0.1,\dyy+0.4);
\draw [->] (\dxx-0.1,\dyy-0.4) -- +(225:1);
\draw [pattern=vertical lines] (4.5,1.5) circle (1.6);
\draw [fill=white] (4.5,1.5) circle (1.4);
\draw [pattern=dots] (4.5,1.5) circle (1.4);
\draw [fill=white] (4.5,1.5) circle (1.05);
\draw [pattern=horizontal lines] (4.5,1.5) circle (1.05);
\draw [fill=white] (4.5,1.5) circle (0.7);
\draw [pattern=north west lines] (4.5,1.5) circle (0.7);
\draw[ultra thin] (1.2,1.5)--(4.5,1.5);
\draw[ultra thin] (1.2,3.1)--(4.5,3.1);
\draw[ultra thin] (1.6,2.9)--(4.5,2.9);
\draw[ultra thin] (2.0,2.55)--(4.5,2.55);
\draw[ultra thin] (2.4,2.2)--(4.5,2.2);
\toplabelline{1.3,1.5}{1.3,3.1}{\NameRmax}
\midlabelline{1.7,1.5}{1.7,2.9}{\NameRii}
\midlabelline{2.1,1.5}{2.1,2.55}{\NameRi}
\midlabelline{2.5,1.5}{2.5,2.2}{\NameR}
\draw [color=white](1.05,-0.2)--(1.05,0);
\node at (4.5,1.5)[circle,fill,inner sep=1pt]{};
\end{tikzpicture}}

\DeclareMathOperator\erf{erf}

\usepackage{color}

\newcommand{\unit}[1]{\ensuremath{\,\mathrm{#1}}}

\newcounter{zoom}
\journal{Discrete and Continuous Models and Applied Computational Science}{\tiny }
\date{}

\begin{document}
\begin{frontmatter}
		%\add{If it easy to do, change numbers in figures 2, 3, and 5 so that you use a dot in between numbers and not commas, e.g. 0,2 $\rightarrow$ 0.2. In (e)ps it should be easy to change inside the file.}
		
		\title{Parallel Algorithm for Numerical Solution of Heat Equation in Complex Cylindrical Domain}
		
		\author[jinr-lit]{Alexander~Ayriyan}\ead{ayriyan@jinr.ru}
		\author[jinr-lit,iep-sas]{J\'an~Bu\v{s}a~Jr.}\ead{busa@jinr.ru}
		
		\address[jinr-lit]{Laboratory of Information Technologies, JINR, Dubna, Russia}
		\address[iep-sas]{Institute of Experimental Physics, Slovak Academy of Sciences, Ko\v{s}ice, Slovakia}
%		\address[ysu]{Deparment of Theoretical Physics, Yerevan State University, Armneia}

		%\cortext[corr]{Основной автор}
		
		\begin{abstract}
        In this article we present a parallel algorithm for simulation of the heat conduction process inside the so-called pulse cryogenic cell. This simulation is important for designing the device for portion injection of working gases into ionization chamber of ion source. The simulation is based on the numerical solving of the quasilinear heat equation with periodic source in a multilayered cylindrical domain. For numerical solution the Alternating Direction Implicit (ADI) method is used. Due to the non-linearity of the heat equation the simple-iteration method has been applied. In order to ensure convergence of the iteration process, the adaptive time-step has been implemented. The parallelization of the calculation has been realized with shared memory application programming interface OpenMP and the performance of the parallel algorithm is in agreement with the case studies in literature.
		\end{abstract}
		
		\begin{keyword}
			Quasilinear heat equation \sep multilayer cylindrical geometrical structure \sep pulse periodic source \sep parallel algorithm \sep thermal gates
			\PACS 07.05.Tp %Computer modeling and simulation
			\sep 02.60.Pn %Numerical optimization
			\sep 02.70.Bf %Finite-difference methods
		\end{keyword}
		
	\end{frontmatter}
	
	%\tableofcontents
	
\section{Introduction}
\label{sec:intro}

The purpose of this work is to develop algorithms for simulation of the heat conduction process inside the so-called pulse cryogenic cell \cite{donets_2012,donets_2015}. Such simulations are important for designing the cell that implements ``the thermal gates'' of a portion injection of working gases into the ionization chamber of a multiply charged ion source \cite{ayriyan_2016_ate}.
While reliable operation of mechanical valves for pulsed injection of gaseous mixtures in the millisecond range at cryogenic temperatures is practically impossible, the use of gas temperature properties at cryogenic temperatures can be a real alternative. Indeed, the vapor pressures of various gases have strong dependency on the temperature \cite{honing_1960} in the interval between temperatures of liquid helium {$4.2\unit{K}$} and liquid nitrogen {$78\unit{K}$} \cite{chelton_1956}, typical temperature terminals in the cryogenic technique used as thermostats with large capacity.

\begin{figure}[H]
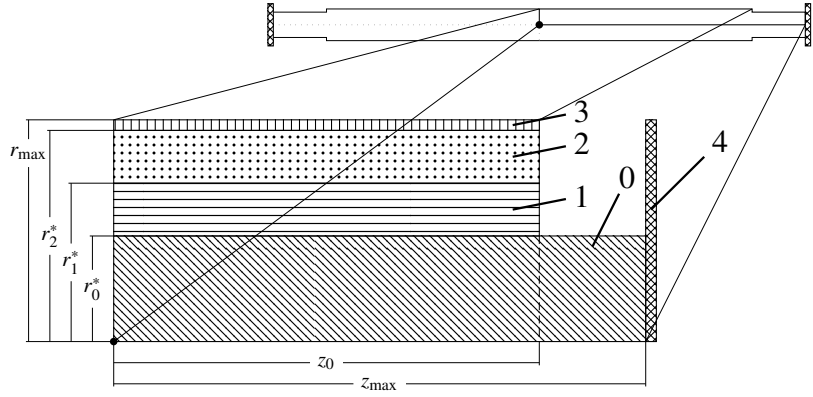

\centering \injectorB{\scriptsize}{0.7}
\caption{A schematic view of the quarter of the cell slice through the axis. The bottom line is the cylinder axis (the symmetry axis, $r = 0$). The cooler (the copper core rod) cools the cell by contact with the temperature terminal (liquid helium). The heater (the conductive layer) heats the cell up by the way of passing of the electrical current. The inner insulator is needed to prevent the electrical current outflow from the heater to the cooler.}
\label{fig:object}
\end{figure}

The cryogenic cell is a multilayer cylinder (see Fig.~\ref{fig:object}) placed inside a vacuum chamber. The thermal process in the cylinder starts by passing an electrical current through its conductive layer. It allows heating the outer layer of the cylinder to upper desired temperature (maximal critical temperature). After switching the current off, the cylinder is let to cool down to lower desired temperature (minimal critical temperature). This process periodically repeats for a given period of time based on the requirements of the experiment. A copper core of the cylinder serves as a cooler for outer layers during and after the heating process. It is connected to a liquid helium temperature terminal. The core is separated by the electrical insulator from the conductive layer. It is made to avoid the flow of the electrical current into the core. The last layer is a thin coating which prevents molecules of working gases from binding to the conductive layer made of graphite.

\section{Initial-Boundary Value Problem}
\label{sec:problem}

\begin{figure}[H]
	\centering
	\includegraphics[width=0.45\textwidth]{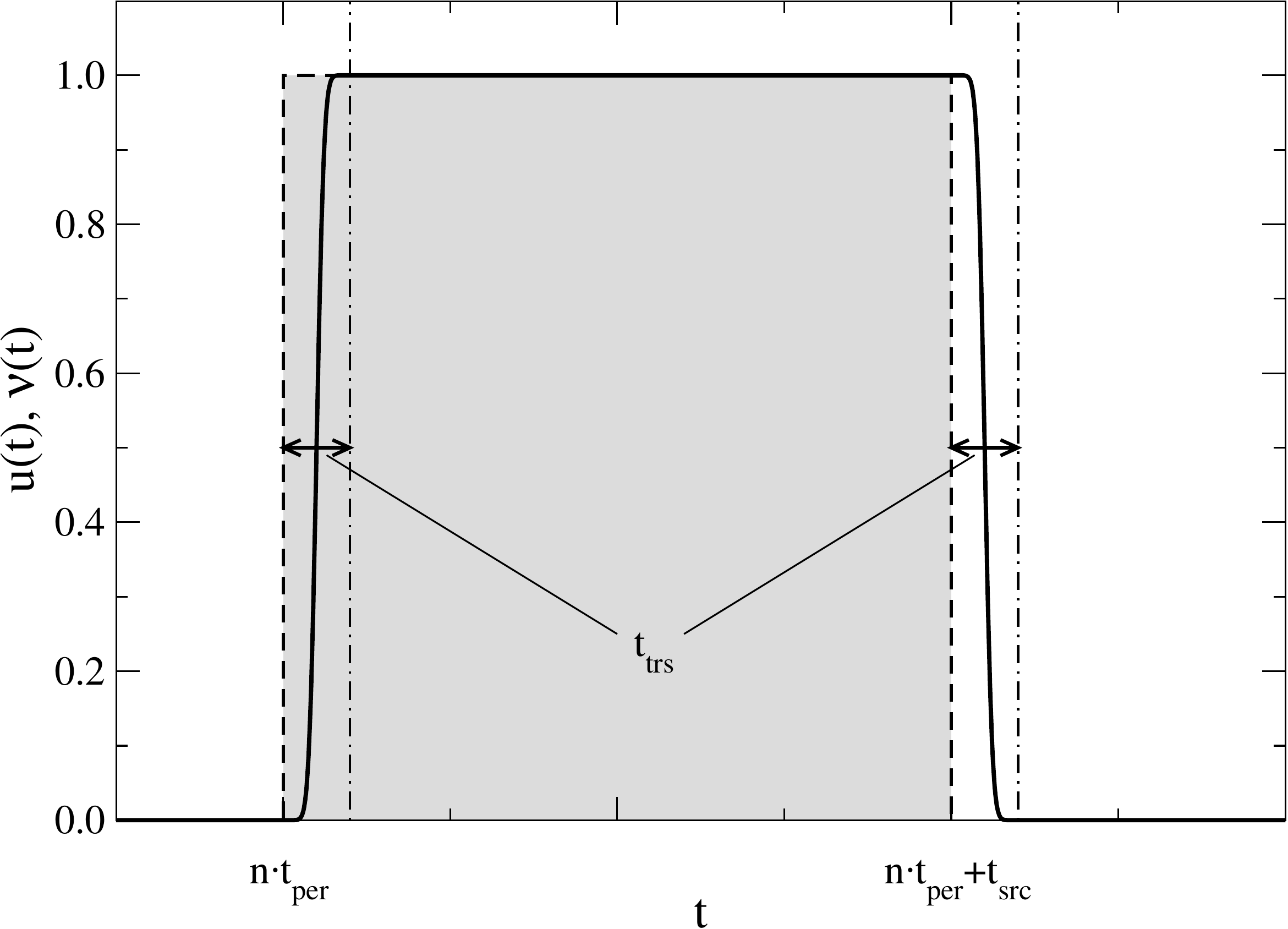}
	\caption{\texttt{Dashed} line: the periodic step function $u(t)$ as a combination of the Heaviside ones. \texttt{Solid} line: the ``realistic'' source function for $\xi=4$ and $\zeta=2$ representing the transient model. The \texttt{dot-dashed} lines show the ends of the transient process at ``turn on'' and ``turn off''. }
	\label{fig:transient}
\end{figure}

Let us consider the heat equation describing thermal evolution in the closed cylindrical domain $\mathbf{\overline{\Omega}} = \lbrace (r, z) \,|\, r\in[0, r_\mathrm{max}(z)], z\in[0, z_\mathrm{max}(r)] \rbrace $:
\begin{equation}
\label{main_eq}
\rho(r) {c_V} (T,r)\frac{\partial T}{\partial t}=\frac{1}{r}
\frac{\partial}{\partial r} \left( r \lambda(T,r) \frac{\partial
	T}{\partial r}\right ) + \frac{\partial}{\partial z} \left( \lambda(T,r) \frac{\partial
	T}{\partial z}\right ) + X(T,t,r; t_{\mathrm{per}}, t_{\mathrm{src}}),
\end{equation}
where the thermal coefficients are the non-linear function of the temperature and they have discontinuities of the first kind along the radial direction at $r^*_m$ ($m=0,1,2,3$, see Fig.~\ref{fig:object}). The source function producing the periodic process of heating can be expressed in the form
\begin{equation}
\label{source}
X(T,t,r; t_{\mathrm{per}}, t_{\mathrm{src}})= \chi(T) \dfrac{I^2(r)}{S_C} p(t; t_{\mathrm{per}}, t_{\mathrm{src}}),
\end{equation}
where $p(t; t_{\mathrm{per}}, t_{\mathrm{src}})$ is the periodic normalized function with parameters $t_{\mathrm{per}}$ (time of period) and $t_{\mathrm{src}}$ (time of heating less or equal $t_{\mathrm{per}}$); $n\in\mathbb{N}_{0}$ is the index of a period; $\chi(T,r)$ is temperature depended specific resistivity with discontinuities at the given values of $r$; $I(r)$ is the electrical current amplitude which has a finite value $I_0$ only in the source layer (see~Fig.~\ref{fig:object}), everywhere else it is zero; $S_C$ is the cross-section of the source layer.
The periodic normalized function $p(t; t_{\mathrm{per}}, t_{\mathrm{src}})$ is given as a rectangular pulse one~\cite{symons_2013}:
\begin{equation}
\label{unit_model}
u(t; t_{\mathrm{per}}, t_{\mathrm{src}}) = \sum_{n=0}^{\infty}\left[ \theta \left(t-n\,t_{\mathrm{per}}\right) - \theta \left(t-n\,t_{\mathrm{per}}-t_{\mathrm{src}}\right) \right],
\end{equation}
where $\theta(t)$ is Heaviside step function~\cite{abramowitz_1972,korn_2013},
however, in order to make the processes of ``turn on'' and ``turn off'' more realistic, the model of the transient process~\eqref{trans_model} (see~Fig.~\ref{fig:transient}) has been implemented:
\begin{equation}
\label{trans_model}
v(t; t_{\mathrm{per}}, t_{\mathrm{src}}, t_{\mathrm{trs}}, \xi, \zeta)  = 
\dfrac{1}{2}\sum_{n=0}^{\infty}
\left[\erf\left(\xi \left(\zeta\dfrac{t-n
	t_{\mathrm{per}}}{t_{\mathrm{trs}}}-1\right)\right) - \erf\left(\xi \left(\zeta\dfrac{t - n\,t_{\mathrm{per}} - t_{\mathrm{src}}}{t_{\mathrm{trs}}}-1\right)\right)\right].
\end{equation}
here $\erf(t)$ is the error function~\cite{abramowitz_1972,korn_2013}.
Such smoothing of the processes of ``turn on'' and ``turn off'' of the source also helps to stabilize numerical simulations.

The initial condition can be set by the assumption that at $t=0$ the cryogenic cell has been already cooled down by the temperature terminal:
\begin{equation}
\label{initcond} 
T(r,z,t=0) = T_0,
\end{equation}
where $T_0 \equiv 4.2$\,K is the temperature of liquid helium.

The boundary conditions can be expressed after the assumption that the temperature flow across the boundary of the domain is zero except for the right side where a connection to the temperature terminal exists (see~Fig.~\ref{fig:object}):
\begin{equation}
\label{boundcond}
\left\{
\begin{array}{l@{\qquad}l}
\displaystyle \frac{\partial T}{\partial \mathbf{n}} =0 & \forall \, (r,z) \in \mathbf{\delta\Omega} \setminus \lbrace (r,z): z=z_\mathrm{max} \rbrace,\\[3mm]
\displaystyle T=T_{0} & \forall \, (r,z) \in \lbrace (r,z): z=z_\mathrm{max} \rbrace,
\end{array}
\right.
\end{equation}
where $\mathbf{\delta\Omega}$ is the boundary of $\mathbf{\Omega}$, $\mathbf{n}$ is the normal vector of the boundary $\mathbf{\delta\Omega}$.
This assumption is motivated by the following statements:
\begin{itemize}
\item the cryonics cell is installed in the vacuum chamber, therefore, there is no convective heat transfer;
\item the working temperature is too low for the appearance of thermal radiation;
\item we neglect the energy for evaporation of gas molecules;
\item there is no temperature flow through the axis $r=0$ due to the axial symmetry.
\end{itemize}

\section{Discretization and Numerical Method}
\label{sec:discretization}

\begin{figure}[H]
	\centering
	\includegraphics[width=0.5\textwidth]{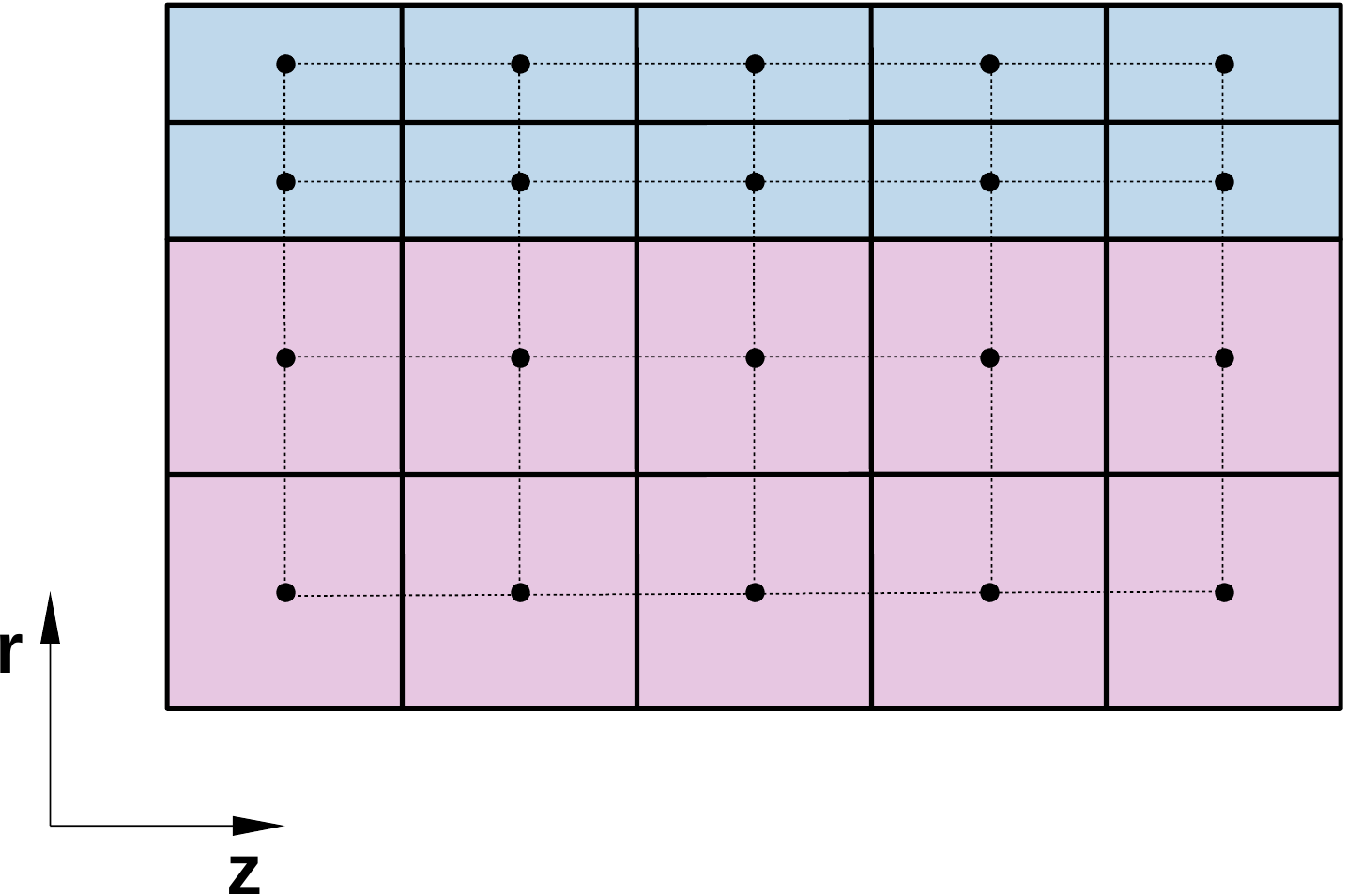}
	\caption{The discretization of the 2D domain. The non-uniform grid is shifted in order to have points at the centers of boxes. Different layers (materials) have different step size (they are differently colored).}
	\label{fig:clattice}
\end{figure}

Numerical solution of Eq.~(\ref{main_eq}) can be obtained by using a shifted non-uniform grid (see~Fig.~\ref{fig:clattice}):
\begin{eqnarray}
\label{grid}
\overline{\omega}& = \lbrace (t,r,z)\left|\right. & 0 \leq t < \infty,\quad t_{k+1} = t_k + \tau_{k+1},\quad k \in \mathbb{N}_{0};\nonumber\\
& & 0.5h_1 \leq r \leq r_{\mathrm{max}}-0.5h_{N_j-1},\quad r_{i+1} = r_i+h_{i+1},\quad i = 0,\ldots, N_j-1;\\
& & 0.5\eta_1 \leq z \leq z_{\mathrm{max}}-0.5\eta_{M_i-1},\quad z_{j+1} = z_j+\eta_{j+1},\quad j = 0, \ldots, M_i-1\nonumber
\rbrace.
\end{eqnarray}

It seems usual to make uniform steps in the subdomain corresponding one layer. The shifted grid has no points at the boundary of discontinuity. One can also use a special grid -- the grid with points at the boundary of materials, in this case one has to take care of approximation of the thermal coefficients and the source function at the boundary.

The initial boundary value problem Eqs.~\eqref{main_eq}--\eqref{boundcond} has been approximated on the grid~\eqref{grid} by the alternating direction implicit (ADI) schemes \cite{peaceman_1955, yanenko_1967, samarskii_1999, samarski_2001}:
\begin{eqnarray}
\label{scheme_eq_1}
\overline{\rho}_{i,j}\,{\overline{c}_V}_{i,j}\frac{\overline{T}_{i,j}-T_{i,j}}{0.5\tau}=
{\textrm{\Large{$\Lambda_r$}}}\left[\right.\overline{T}_{i,j}\left.\right]+
{\textrm{\Large{$\Lambda_z$}}}\left[\right.T_{i,j}\left.\right]+
\overline{X}_{i,j},\\[4mm]
\label{scheme_eq_2}
\overline{\rho}_{i,j}\,{\overline{c}_V}_{i,j}\frac{\widehat{T}_{i,j}-\overline{T}_{i,j}}{0.5\tau}=
{\textrm{\Large{$\Lambda_r$}}}\left[\right.\overline{T}_{i,j}\left.\right]+
{\textrm{\Large{$\Lambda_z$}}}\left[\right.\widehat{T}_{i,j}\left.\right]+
\overline{X}_{i,j},
\end{eqnarray}
where $\widehat{T}_{i,j}$ is the temperature distribution on the next time layer, $\overline{T}_{i,j}$ is the temperature distribution on the half layer (in between the next and current time layers), $T_{i,j}$ is the value on the current time layer and $\tau$ is the time step, $\overline{\rho}_{i,j}=\rho(\overline{T}_{i,j})$, ${\overline{c}_V}_{i,j}=c_V(\overline{T}_{i,j}),$ and $\overline{X}_{i,j}=X(\overline{T}_{i,j})$.

The spatial finite difference operators in Eqs.~\eqref{scheme_eq_1} and \eqref{scheme_eq_2}  are:
\begin{equation}
\vspace{2mm}
\label{Lambda_r}
{\textrm{\Large{$\Lambda_r$}}}\left[{T}_{i,j}\right]=
\frac{1}{r_i}\frac{1}{\hbar_{i}}\left[r_{i+\frac{1}{2}} \lambda_{i+\frac{1}{2},j} 
\frac{{T}_{i+1,j}-{T}_{i,j}}{h_{i+1}}-r_{i-\frac{1}{2}} \lambda_{i-\frac{1}{2},j}
\frac{{T}_{i,j}-{T}_{i-1,j}}{h_{i}}\right],
\end{equation}
%\vspace{2mm}
\def\bareta{\eta\kern-0.8ex\lower0.2ex\hbox{\vrule height0.4pt width0.8ex}\kern0.1ex}
\begin{equation}
\label{Lambda_z}
{\textrm{\Large{$\Lambda_z$}}}\left[\right.T_{i,j}\left.\right]=
\frac{1}{\bareta_j}\left[\lambda_{i,j+\frac{1}{2}} 
\frac{T_{i,j+1}-T_{i,j}}{\eta_{j+1}}-\lambda_{i,j-\frac{1}{2}}
\frac{T_{i,j}-T_{i,j-1}}{\eta_{j}}\right],
\vspace{2mm}
\end{equation}
where $i = 1,\ldots,N_j-1$, $j = 1,\ldots, M_i-1$, $h_{i} = r_{i}-r_{i-1}$, $\eta_{j} = z_{j}-z_{j-1}$, $\hbar_{i}=\left(h_{i+1}+h_{i}\right)/2$, $\bareta_j=\left(\eta_{j+1}+\eta_{j}\right)/2$, $\displaystyle \lambda_{i,j}=\lambda_m(T_{i,j})$, $\displaystyle {c_V}_{i,j}={c_V}_m(T_{i,j})$, $\displaystyle X_{i,j}=X_m(T_{i,j})$, $\displaystyle r_{i\pm 1/2}=(r_i + r_{i\pm1})/2$, \linebreak $\displaystyle \lambda_{i\pm 1/2,j}=\lambda_m(T_{i,j}+T_{i\pm1,j})/2$, $\displaystyle \lambda_{i,j\pm 1/2}=\lambda_m(T_{i,j}+T_{i,j\pm1})/2$.

Due to the non-linearity of Eqs.~\eqref{scheme_eq_1}--\eqref{scheme_eq_2} (when the thermal coefficients and the source function depend on temperature) the simple-iteration method has been applied for calculation of the sought-for function on the half and next time layers. The recursive forms for \eqref{scheme_eq_1}--\eqref{scheme_eq_2} are expressed as follows \cite{samarskii_1999, samarski_2001, kalitkin_2013}:
\begin{eqnarray}
	\label{iter_scheme_eq_1}
	\rho^{s}_{i,j}\,{c_V}^{s}_{i,j}\frac{T^{s+1}_{i,j}-T_{i,j}}{0.5\tau}=
	\frac{1}{r_i}\frac{1}{\hbar_{i}}\left[r_{i+\frac{1}{2}} \lambda^{s}_{i+\frac{1}{2},j} 
	\frac{T^{s+1}_{i+1,j}-T^{s+1}_{i,j}}{h_{i+1}}-r_{i-\frac{1}{2}} \lambda^{s}_{i-\frac{1}{2},j}
	\frac{T^{s+1}_{i,j}-T^{s+1}_{i-1,j}}{h_{i}}\right]+
	{\textrm{\Large{$\Lambda_z$}}}\left[\right.T_{i,j}\left.\right]+
	X^{s}_{i,j},
\end{eqnarray}
here it is supposed that when $s\rightarrow \infty,$ then $T^{s} \rightarrow \overline{T}$, $\rho^{s} \rightarrow \overline{\rho}$, ${c_V}^{s} \rightarrow {\overline{c}_V}$, $\lambda^{s} \rightarrow \overline{\lambda},$ and  $X^{s} \rightarrow \overline{X}$. The iteration process starts with the initial condition that $T^{s=0}_{i,j}=T_{i,j}$, and stops after fulfilling the following criteria
\begin{equation}
\vspace{2mm}
\label{iter_cond}
||T^{s+1}-T^{s}||_{C} = \max\limits_{\overline{\omega}}\|T^{s+1}-T^{s}\|< \varepsilon.
\end{equation}

The values of the sought-for function on next time-layer are obtained as this
\begin{eqnarray}
	\label{iter_scheme_eq_2}
	\overline{\rho}_{i,j}\,{\overline{c}_V}_{i,j}\frac{T^{s+1}_{i,j}-\overline{T}_{i,j}}{0.5\tau}=
	{\textrm{\Large{$\Lambda_r$}}}\left[\right.\overline{T}_{i,j}\left.\right]+
	\frac{1}{\bareta_j}\left[\lambda^{s}_{i,j+\frac{1}{2}} 
	\frac{T^{s+1}_{i,j+1}-T^{s+1}_{i,j}}{\eta_{j+1}}-\lambda^{s}_{i,j-\frac{1}{2}}
	\frac{T^{s+1}_{i,j}-T^{s+1}_{i,j-1}}{\eta_{j}}\right]+
	\overline{X}_{i,j},
\end{eqnarray}
with the initial condition $T^{s=0}_{i,j}=\overline{T}_{i,j}$. Same as before $T^{s\rightarrow \infty} \rightarrow \widehat{T}$ and $\lambda^{s\rightarrow \infty} \rightarrow \hat{\lambda}$. This iteration process stops after fulfilling the same criteria~\eqref{iter_cond}. The systems of linear algebraic equations \eqref{iter_scheme_eq_1} and \eqref{iter_scheme_eq_2} are solved by the Thomas method~\cite{thomas_1949,samarski_1995,kalitkin_2013}.

In order to ensure convergence of the iteration process \eqref{iter_scheme_eq_1}--\eqref{iter_scheme_eq_2}, the adaptive time-step has been implemented. If the counter of iterations $s$ exceeds some maximal value, meaning that the process does not converge fast enough, the time-step $\tau$ is divided by 2 and the iteration process is restarted (back to the Eq.~\eqref{iter_scheme_eq_1}).

\section{Parallel Algorithm}
\label{sec:algorithm}

\begin{figure}
\begin{center}
  \includegraphics[width=0.81\textwidth]{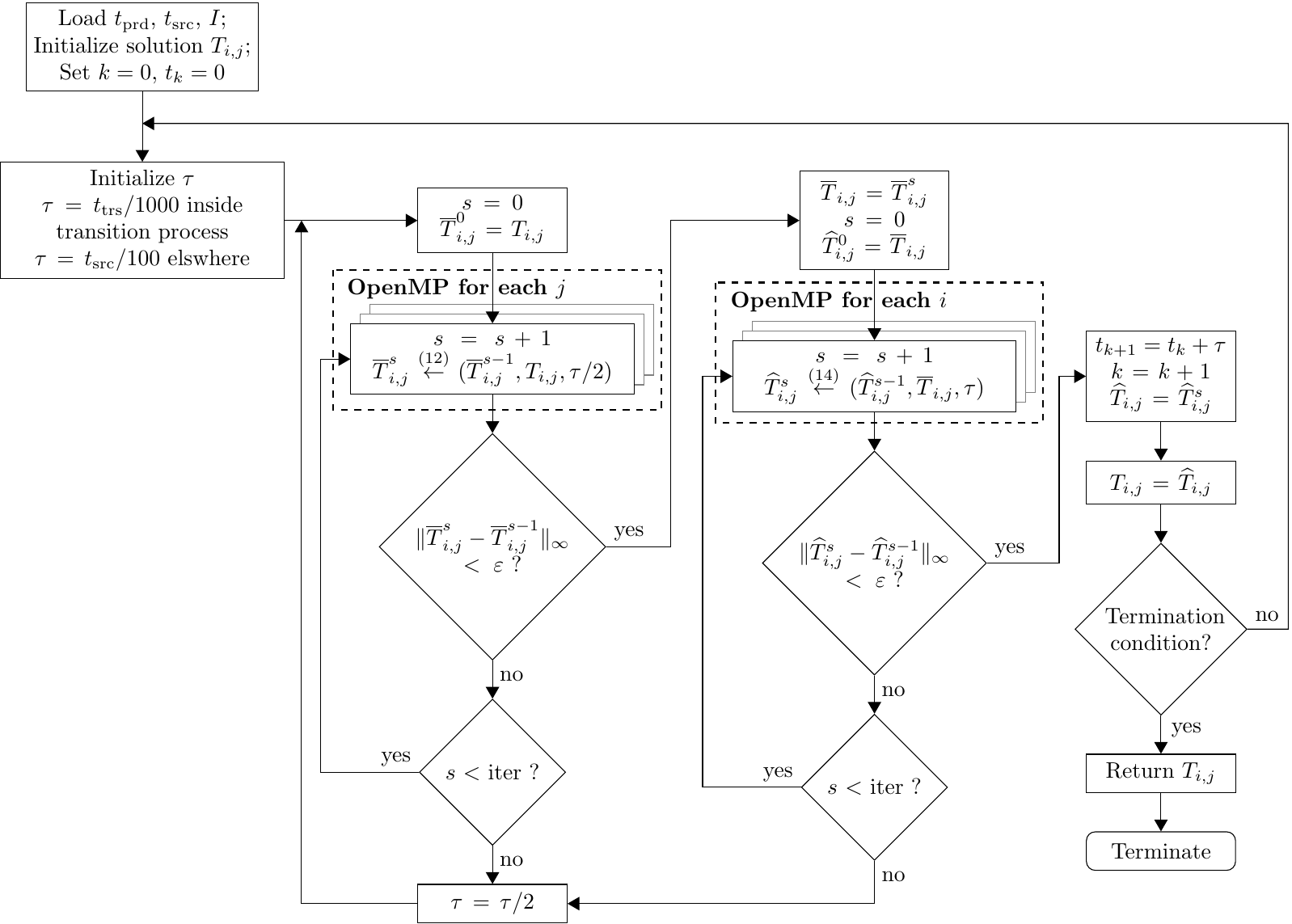}
\end{center}
\caption{Schematics of the algorithm with parts running using OpenMP highlighted by the dashed line.}
\label{fig:flwchrtAlgOpenMP}
\end{figure}

Main computational complexity comes from repetitive calculation of Eqs.~\eqref{iter_scheme_eq_1} and \eqref{iter_scheme_eq_2} across each time layer. A solution of the first of them is needed as a start for solving the second one and it is again used as a start for calculation of the sought-for function on the next time layer. Due to rather low complexity of solving of one system of linear equations (\eqref{iter_scheme_eq_1} or \eqref{iter_scheme_eq_2}) it is better to use parallelization based on shared memory since for distributed memory parallelization the cost of data transfer would be too high. Therefore, we opted for OpenMP \cite{dagum_1998openmp,omp_2018}.

In Fig.~\ref{fig:flwchrtAlgOpenMP} one can see the flowchart of the algorithm. After initializing the solution -- setting $t_\mathrm{prd},$ $t_\mathrm{src},$ $I$, and $t_0=0,$ the program repeats iterations until the requested time is reached. In one step of evolution it first initializes the estimate of the time step $\tau$ as being equal $t_\mathrm{trs}/1000$ if we are in the transition process or $t_\mathrm{src}/100$ elsewhere. After this it starts the iteration process in order to obtain the solution $\overline{T}_{i,j}$ at the time $t_{k}+\tau/2$ using \eqref{iter_scheme_eq_1}. If the obtained solution is precise enough, it alternates direction and continues by the way of \eqref{iter_scheme_eq_2} to the solution $\widehat{T}_{i,j}$ at the time $t_{k}+\tau$. If this solution is again precise enough, the algorithm sets the total time to $t_k=t_{k-1}+\tau$ and the actual solution $T_{i,j}=\widehat{T}_{i,j}.$
If in any of the previous tests the number of iterations exceeds the number of maximum iterations (iter -- see Fig.~\ref{fig:flwchrtAlgOpenMP}), then the time-step $\tau$ is divided by 2 and the calculation returns to the beginning of the evolution step.

The whole algorithm terminates when the total time $t$ reaches the desired value or when we realize that we have entered into the periodic process (temperature changes periodically depending on switching the current in conductive layer on and off).

%\begin{figure}%[H]
%	\begin{center}
%		\includegraphics[width=0.32\textwidth]{time.pdf}\includegraphics[width=0.32\textwidth]{speedup.pdf}\includegraphics[width=0.34\textwidth]{efficiency.pdf}
%	\end{center}
%	\caption{The experimental performance of the algorithm for $t_{\mathrm{per}}=0.1$ and $t_{\mathrm{src}}=0.01$. \textit{Left} panel: calculation time in hours. \textit{Central} panel: speed up of calculations. \textit{Right} panel: the efficiency of the parallelization in percents.}
%\end{figure}

\section{Numerical Results}
\label{sec:results}

\begin{figure}[p]
	\begin{center}
		\includegraphics[height=0.24\textheight]{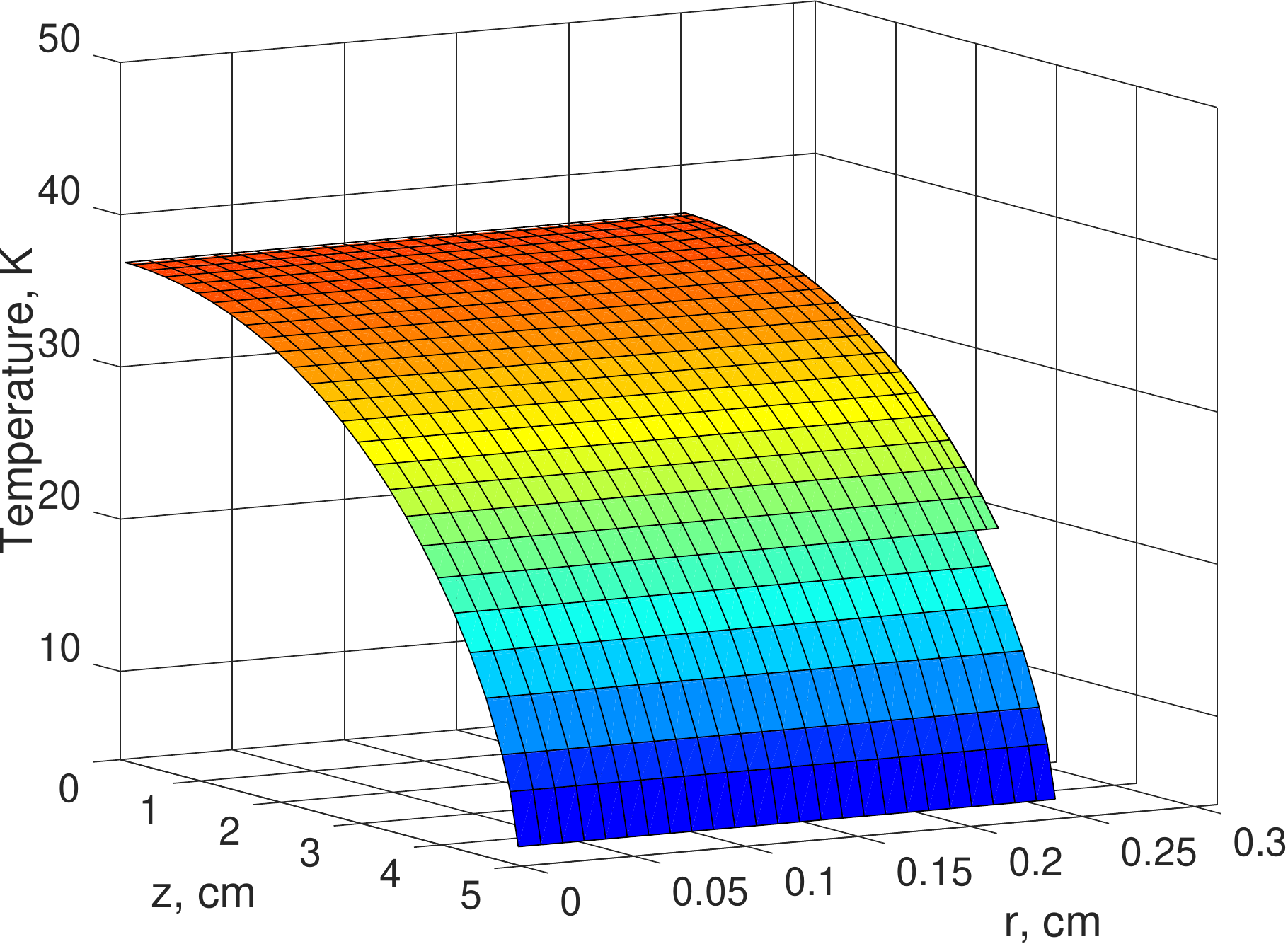} \hspace*{2.5mm}
		\includegraphics[height=0.24\textheight]{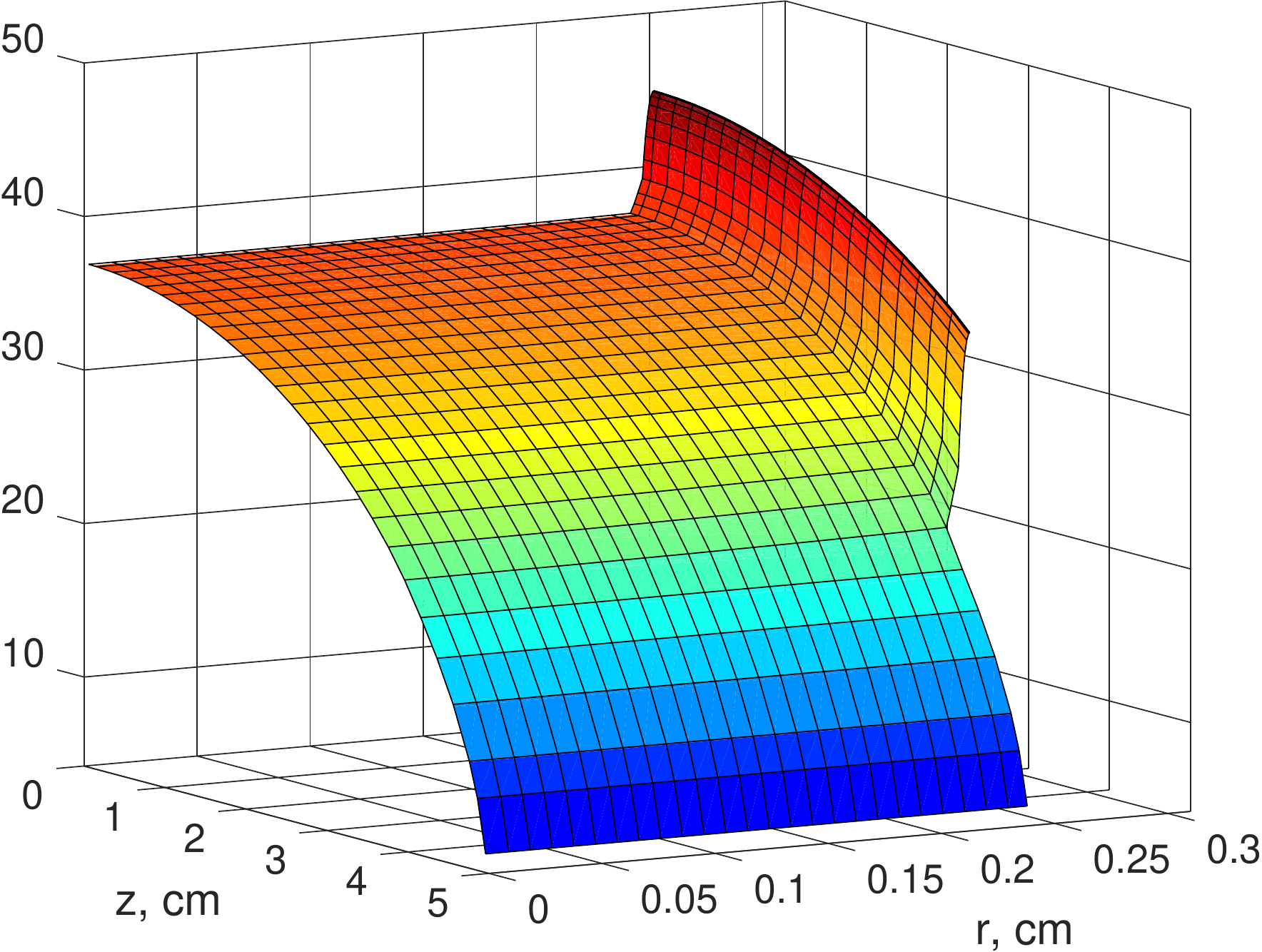} \hspace*{2.5mm}
		\raisebox{5mm}{\includegraphics[height=0.2\textheight]{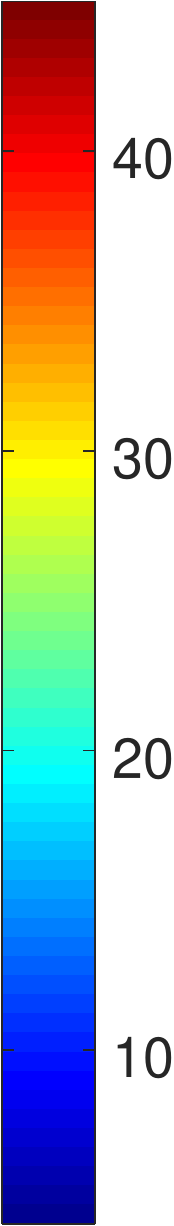}}
	\end{center}
	\caption{Temperature fields when the periodic temperature regime is achieved. \textit{Left} panel:  the temperature field at the moment before turning on the source. \textit{Right} panel: the temperature field at the moment when the source is just turned off.}
	\label{fig:fields}
\end{figure}

\begin{figure}[p]
		\begin{center}
			\includegraphics[height=0.205\textheight]{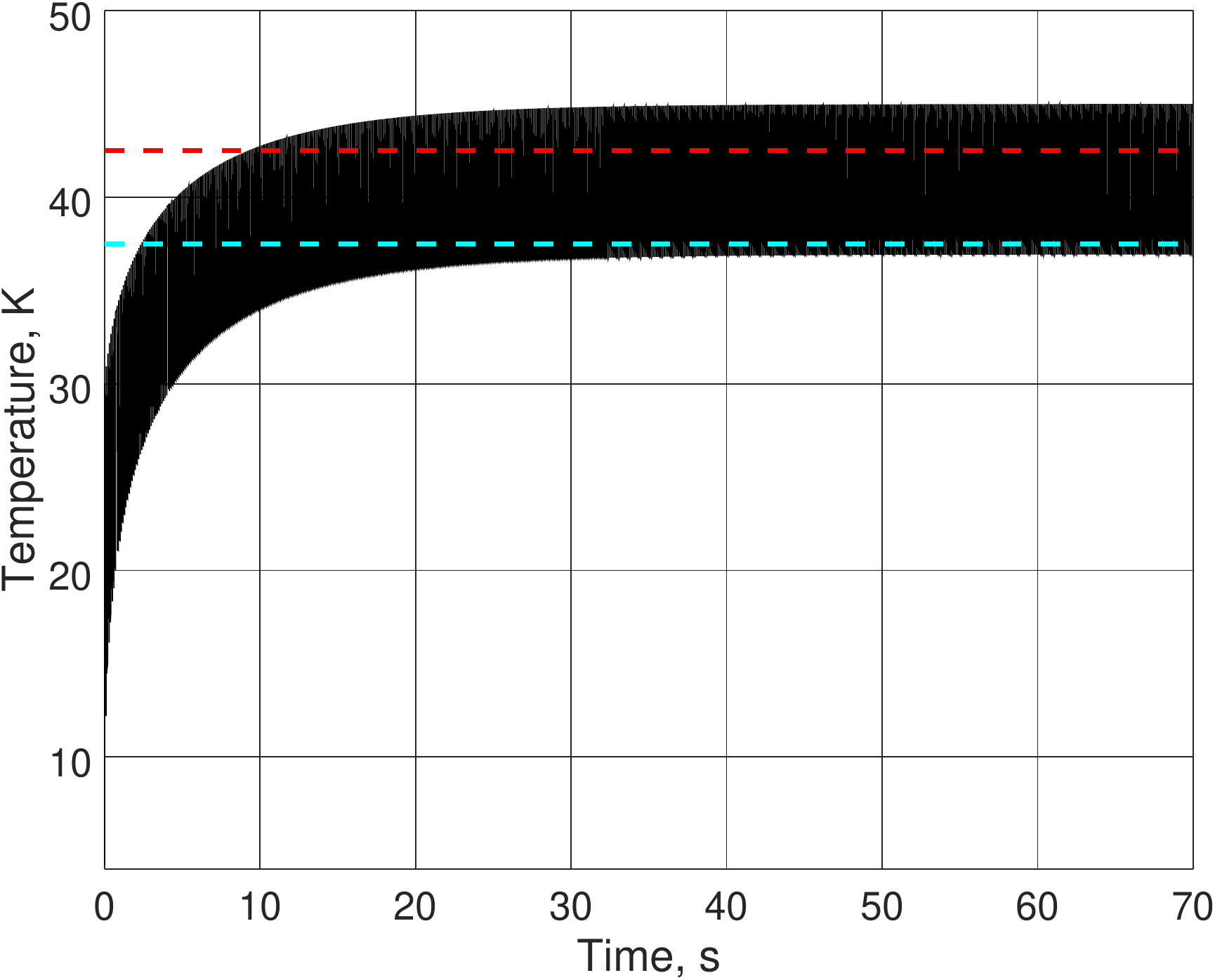}\includegraphics[height=0.205\textheight]{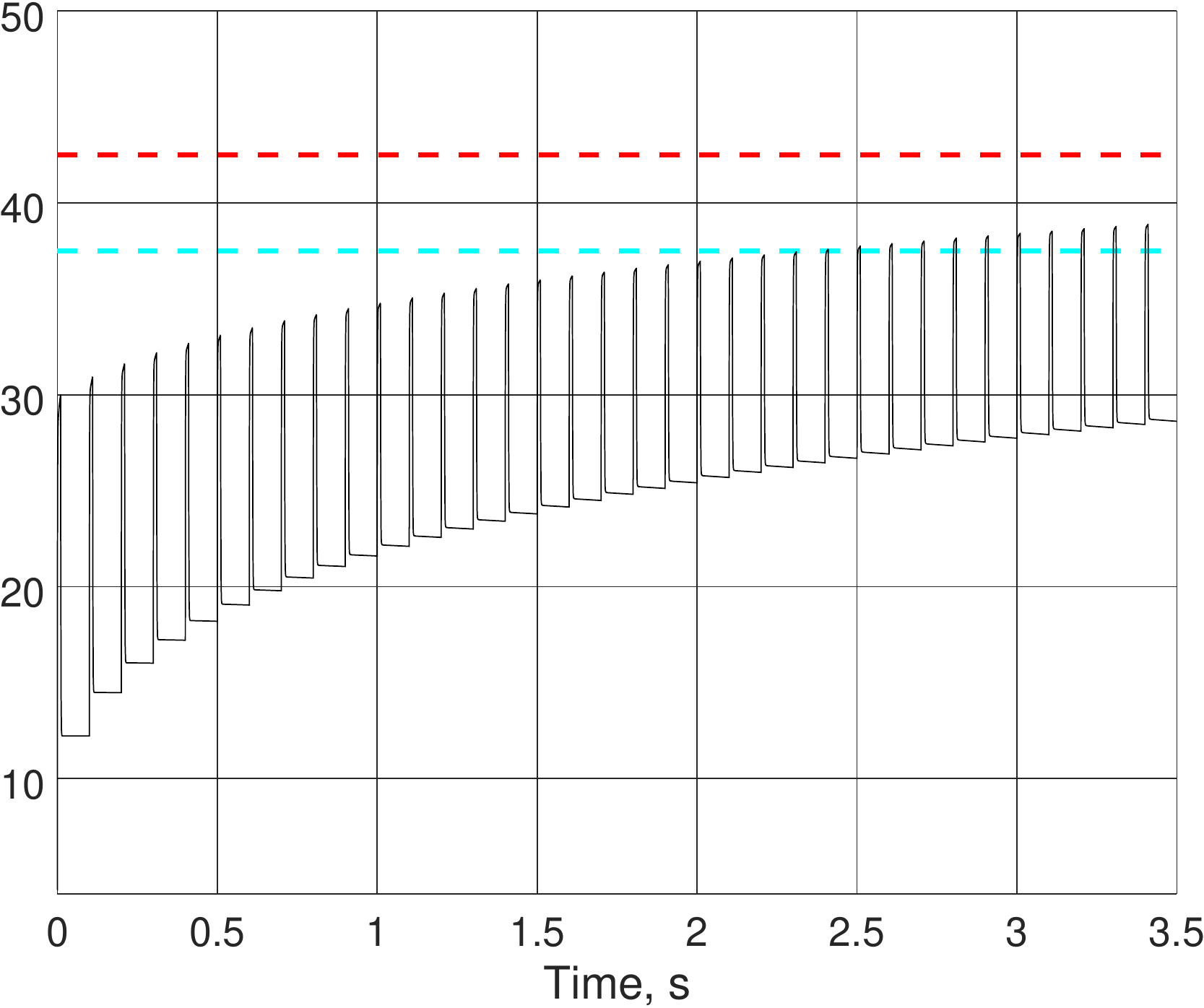}\includegraphics[height=0.205\textheight]{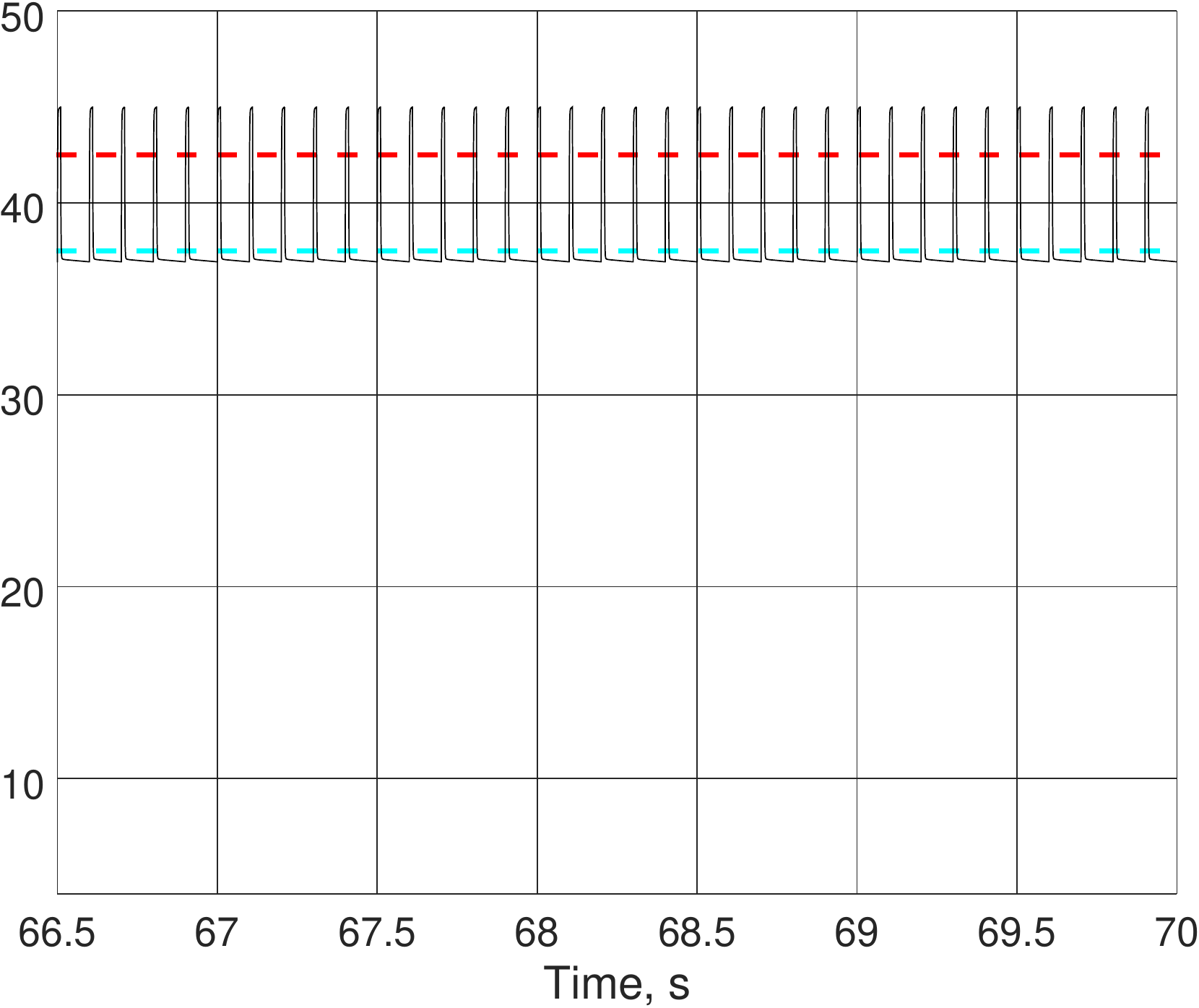}
		\end{center}
		\caption{Temperature evolution on the surface at the $z=0$ (black line). Red colored upper dashed line represents the maximal critical temperature (42\,K) and blue colored lower dashed line represents the minimal critical temperature (37\,K). \textit{Left} panel: Temperature dependence on the time from $0$ up to $70$~sec. \textit{Central} panel: the beginning of the evolution (the first $3.5$~sec.). \textit{Right} panel: the last $3.5$~sec. of the evolution when the periodic steady temperature regime is achieved.}
		\label{fig:tempEvolution}
	\end{figure}

The calculations have been performed on the HybriLIT computational platform~\cite{hlit, adam_2018} using the Skylake processor Intel Xeon Gold 6154 \cite{skylake_6154} containing 18 CPU cores providing two threads per core (36 threads in total) under OS Scientific Linux 7.5 (Nitrogen) \cite{os_sl_7_5}.
As an example we have studied the case when $t_\textrm{per}=10^{-1},$ $t_\textrm{src}=10^{-2},$ $t_\textrm{trs}=10^{-4},$ and $I_0=0.5742$. 
%These values of parameters are the solution of the optimization problem (see details in \cite{ayriyan_ppnl_2019}), they were chosen to realize the ``thermal gates'' for maximal and minimal critical temperatures 42\,K and 37\,K, respectively.
These values of parameters were chosen to realize the ``thermal gates'' for maximal and minimal critical temperatures 42\,K and 37\,K, respectively~\cite{ayriyan_ppnl_2019}. The cell size was selected as follows: $z_0=4$, $z_{\max}=5$, $r_0^*=0.24$, $r_1^*=0.245$, $r_2^*=0.25$, $r_{\max}=0.2501$. The whole domain was split into 100 parts along the $z$ axis at the first layer (core) and to 80 in the other layers. Along the $r$ axis, individual layers were divided (starting from the core) into 800, 200, 200, and 10 parts respectively. This discretization of the domain was chosen in order not to split the domain into too many parts and at the same time have enough information about the solution. It is noteworthy that we have much more steps in the radial direction. It is so because we expect the flow in-between the layers to be more active than the relaxation towards the terminal. Nevertheless, in our experiments we have also densified the grid to see the impact of the grid on both the precision of the solution and the calculation time.
The results are given in Fig.~\ref{fig:fields}, it shows the temperature fields at differen moment of evolution inside the required steady periodic temperature regime. The field is shown in the whole cylindrical domain right before turning on the source (at the left panel). The field is in a state of maximum relaxation in that moment. At the right side the temperature filed is shown in the moment when source is turning off.
The results demonstrate that temperature significantly changes only in outer layers, the core rode practically is not heated, it works as thermostat.
The evolution of the temperature at the surface of the cryogenic cell at the $z=0$ is given in Fig.~\ref{fig:tempEvolution}, it is evident that is takes time to achived the required steady periodic temperature regime.

\begin{figure}[p]
	\begin{center}
		\includegraphics[width=0.32\textwidth]{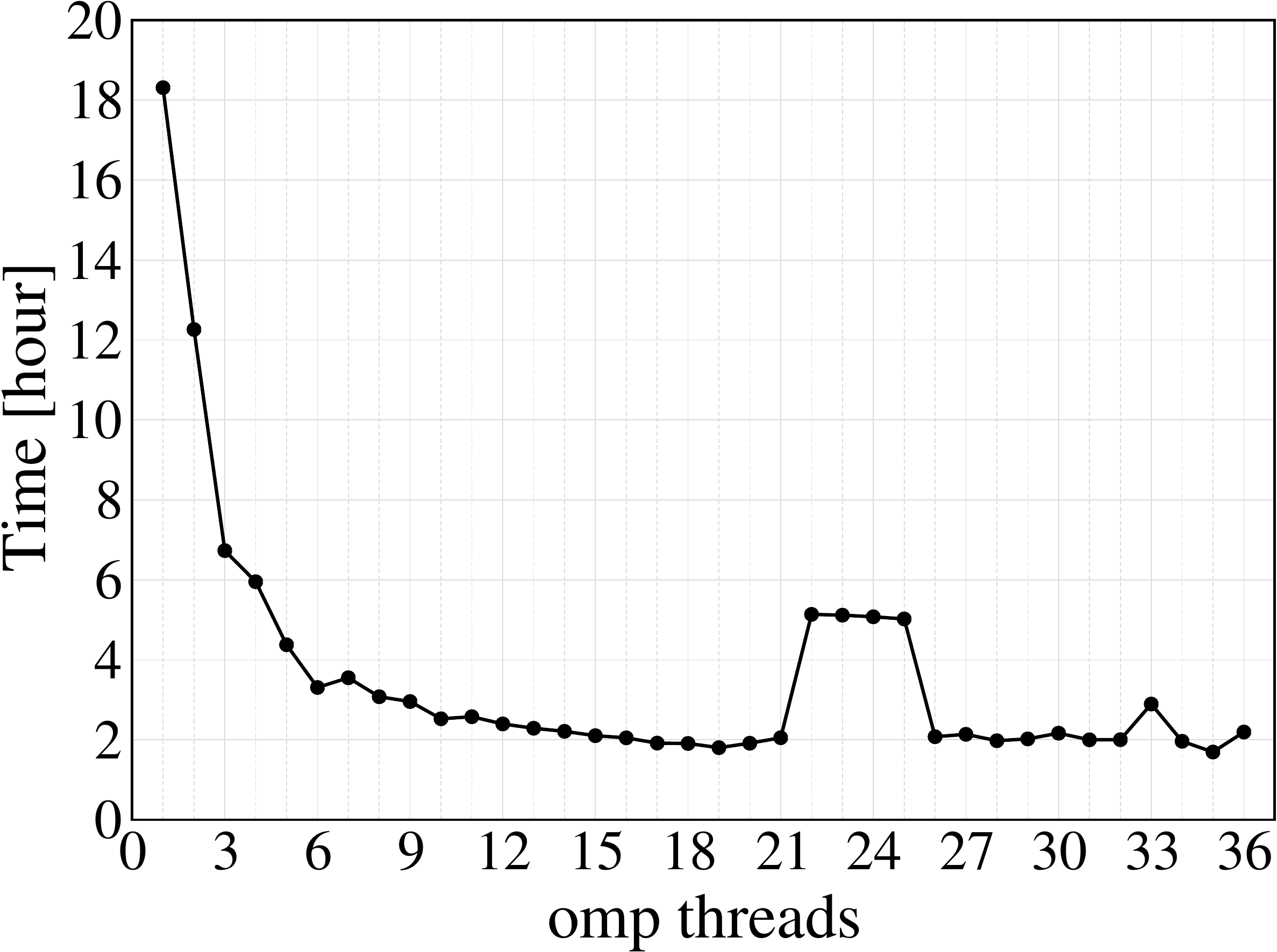}\hspace{1mm}\includegraphics[width=0.32\textwidth]{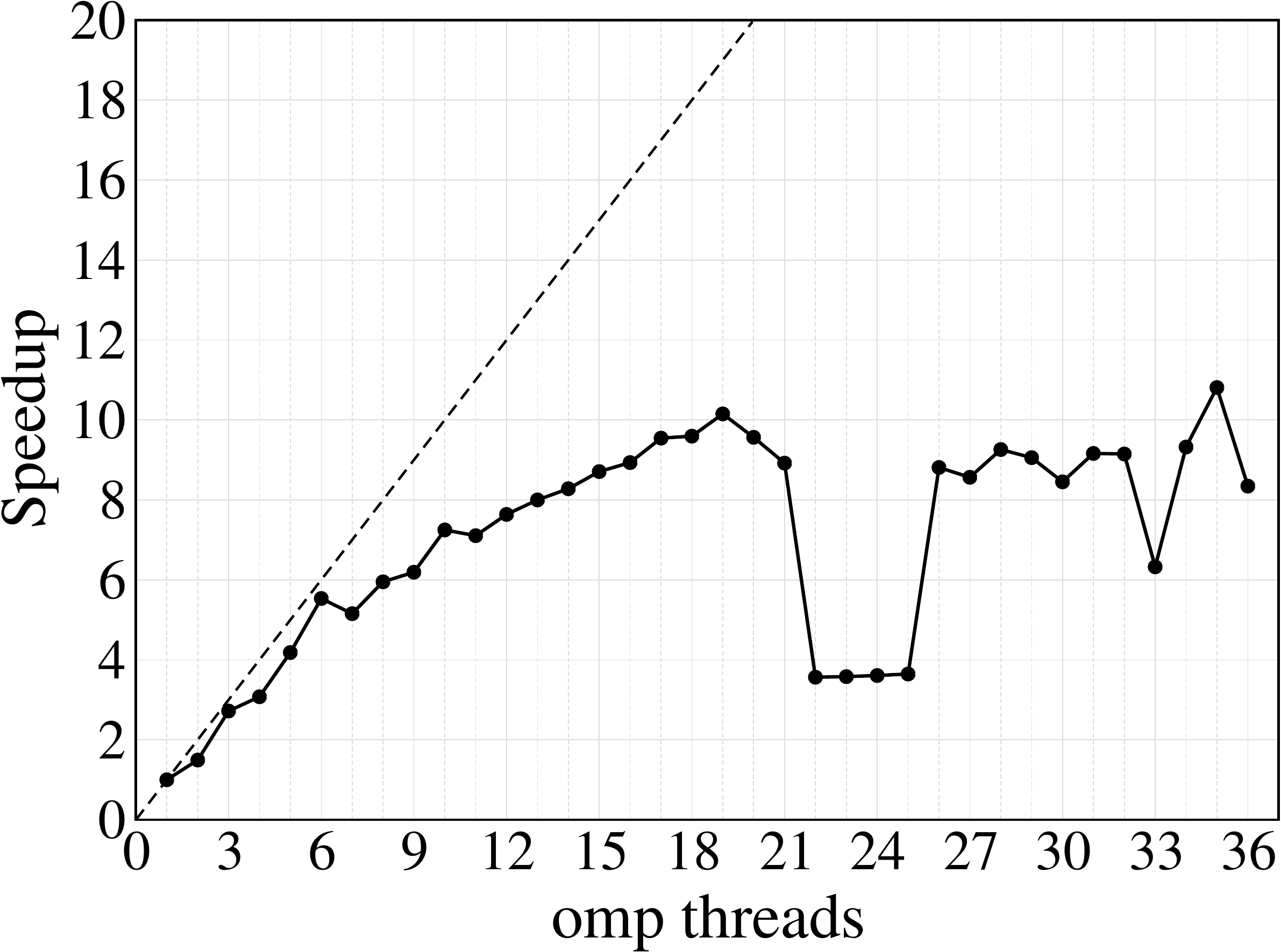}\hspace{1mm}\includegraphics[width=0.33\textwidth]{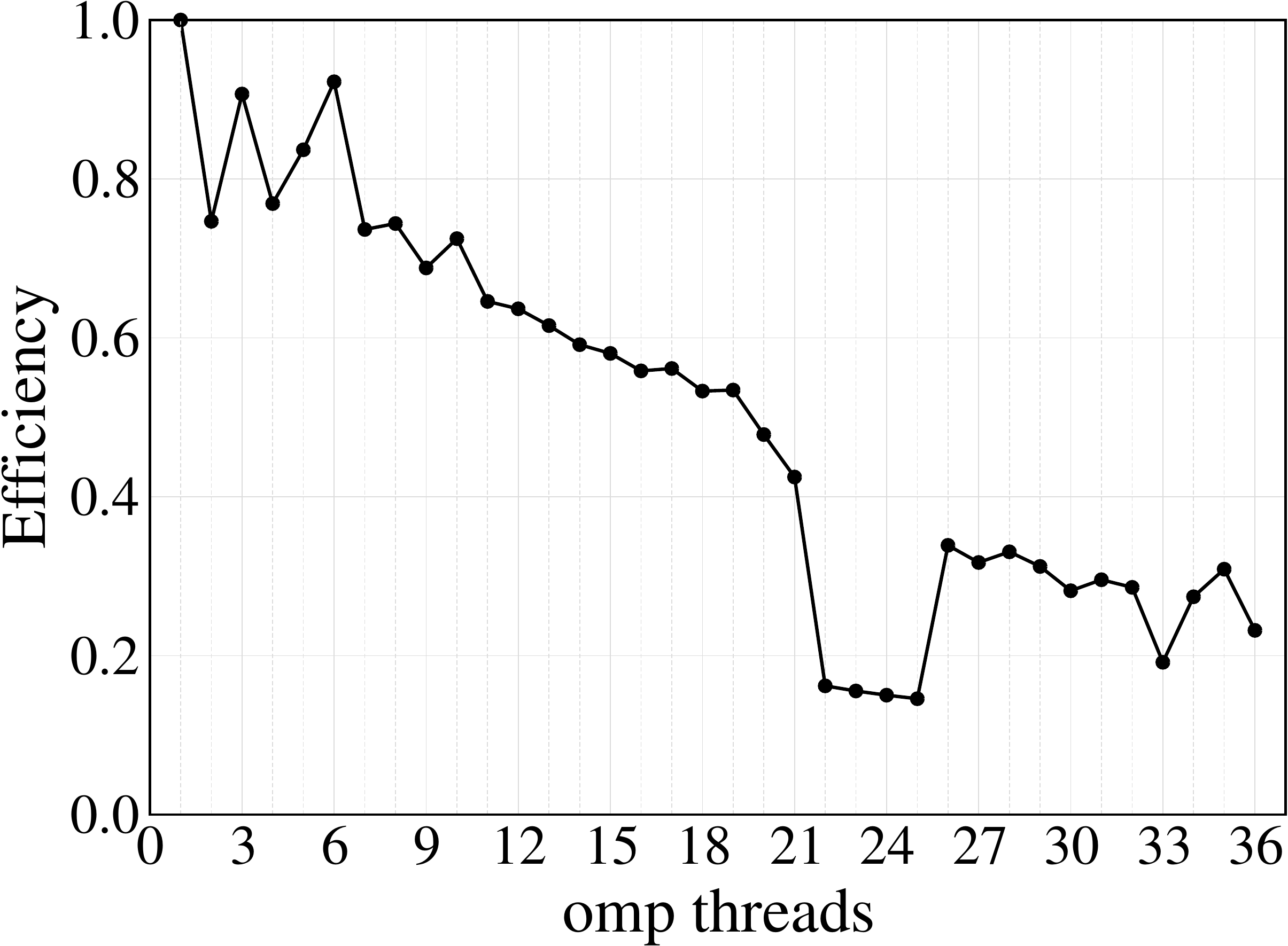}
	\end{center}
	%\caption{The experimental performance of the algorithm for $t_{\mathrm{per}}=0.1$ and $t_{\mathrm{src}}=0.05$. \textit{Left} panel: calculation time in hours. \textit{Central} panel: speed up of calculations. \textit{Right} panel: the efficiency of the parallelization in percents.}
	\caption{The experimental performance of the algorithm. \textit{Left} panel: the calculation time in hours. \textit{Central} panel: the speed up of calculations. \textit{Right} panel: the efficiency of parallelization.}
	\label{fig:omp_speedup}
\end{figure}

\section{Summary and Conclusions}
\label{sec:concl}

We have developed the algorithm for numerical simulations of heat evolution inside the multilayered cylindrical domain with the periodic source. One can see from Fig.~\ref{fig:tempEvolution} that the required periodical temperature regime cannot be obtained immediately, it has a setup mode which is much longer than one period, thus it has to be taken into account when designing the pulse cryogenic cell. The simulations show the possibility of the realization of ``thermal gates'' for a particular set of parameters. The algorithm has been integrated to the hybrid algorithm MPI+OpenMP for solving the optimization problem of the heat source characteristics ($t_{\mathrm{per}}$, $t_{\mathrm{src}},$ and $I_0$) of the pulse cryogenic cell~\cite{ayriyan_ppnl_2019}.

The performance of the parallel algorithm (see Fig.~\ref{fig:omp_speedup}) is in agreement with the case studies in literature, e.g., \cite{tokareva_2016, schuster_2018}. As it is shown in the picture that the saturation calculation time been achieved at 18 threads, after acting the hyper-threading the speedup stops. Thus there is no reason to involve in calculations more than this number of threads for the considered problem with the given grid size ($1210\times100$).

%With use of the described algorithm the optimization problem for the characteristics of the thermal source of a cryogenic cell has been solved by hybrid approach for parallel calculation MPI+OpenMP (see~).

\section*{Acknowledgements}
Authors thank Dr.~H.~Grigorian, Dr.~E.~E.~Donets and E.~Ayryan for fruitful discussions and help, it has allowed to improve the article. The work has been partially supported by the Russian Foundation for Basic Research under the projects \#18-51-18005 and \#19-01-00645. Computations have been held on the basis of the HybriLIT heterogeneous computing platform (LIT, JINR).

\bibliographystyle{unsrtnat}
\biboptions{sort&compress}
\bibliography{heq}

\end{document}